\documentclass[english,floatfix,twocolumn,prl,aps]{revtex4}
\usepackage[T1]{fontenc}
\usepackage[latin1]{inputenc}
\usepackage{graphicx}
\usepackage{amssymb}
\usepackage{amsmath}
\usepackage{bm}
\usepackage{babel}
\usepackage{color}
\makeatother

\begin{document}

\title{Raman Spectroscopy of Graphene Edges}

\author{C. Casiraghi$^1$, A. Hartschuh$^2$, H. Qian$^2$, S. Piscanec$^1$, C. Georgi$^2$,\\K. S. Novoselov$^3$, D. M. Basko$^4$, A. C. Ferrari$^1$}
\affiliation{$^1$Engineering Department, Cambridge University, Cambridge, UK\\
$^2$Chemistry and Biochemistry Department and CeNS,
Ludwig-Maximilians- University of Munich, Germany\\
$^3$Department of Physics and Astronomy, Manchester University, UK\\
$^4$Laboratoire de Physique et Mod\'elisation des Mileux Condens\'es,
Universit\'e Joseph Fourier and CNRS, Grenoble, France\\}

\begin{abstract}

Graphene edges are of particular interest, since their
chirality determines the electronic properties. Here we present a detailed Raman investigation of
graphene flakes with well defined edges oriented at different crystallographic directions. The position, width and intensity of G and D peaks at the edges are studied as a function of the incident light polarization. The D-band is strongest for light polarized parallel to the edge
and minimum for perpendicular orientation. Raman mapping shows that the D peak is localized in proximity of the edge. The D to G ratio does not always show a significant dependence on edge orientation. Thus, even though edges can appear macroscopically smooth and oriented at well defined angles, they are not necessarily microscopically ordered.

\end{abstract}

\maketitle

Graphene is the latest carbon allotrope to be discovered, and it is now at
the center of a significant experimental and theoretical research effort\cite{Nov306(2004),GeimRevNM, Nov438(2005),CastroNetoRev,charlier,Zhang438(2005)}. In particular,
near-ballistic transport at room temperature and high carrier mobilities (between 3000 and 200000
cm$^{2}$/Vs)\cite{Nov438(2005),Zhang438(2005),Nov315(2007),MorozovNov(2007),andrei,kimmob}make it a potential material for
nanoelectronics\cite{Han, Chen,Zhang86,Lemme}, especially for high frequency applications.

Graphene layers can be readily identified in terms of number and orientation by elastic and inelastic light scattering, such as Raman\cite{Casiraghi_condmat_2007, ACFRaman,Pisana,PiscanecPRL,cancado08,Leandro} and Rayleigh spectroscopies\cite{CasiraghiNL,GeimAPL}. Raman spectroscopy also allows monitoring of doping and
defects\cite{Casiraghi_condmat_2007, Pisana,ACFRamanSSC,DasCM}. Once identified, graphene layers
can be processed into nanoribbons by lithography\cite{Nov306(2004),Han,Lemme,Lu75, avourisPE}.

Similar to the case of nanotubes, confinement modifies the electronic structure of graphene,
when cut into nanoribbons\cite{Nakada1996, Fujita1996, Miyamoto59, Wakabayashi59,SonPRL, Pisani,
Nakada1998}. The edges of graphene nanoribbons (GNRs) could in general be a combination of armchair or zigzag regions\cite{Niimi,Kobayashi2005,SolsGuineaCastroNeto,NiimiASS}. If a GNR is uniquely limited by one type of edge, it is defined either as armchair or zigzag\cite{Nakada1996, Fujita1996,NiimiASS}. Edges are also preferred sites for functionalisation with different groups\cite{cervantes}.

Here we show that Raman spectroscopy is a sensitive tool to probe graphene edges. Our results
challenge the suggestion that perfectly armchair or zigzag edges can be easily obtained when
exfoliating graphene, even though they appear to follow defined directions on a large scale.

Single layers are produced by micro-cleavage of graphite. These have areas up to 100$\mu$m$^2$ and show sharp edges with different orientations. Raman spectra are measured with a 100X objective at 514, 633 and 488 nm with a Renishaw micro-Raman spectrometer, having a 1800 grooves/mm grating and spectral resolution of $\sim$3cm$^{-1}$. The polarization of the incident light can be controlled by a Fresnel rhomb. Raman mapping is performed in an inverted confocal microscope at 633nm. The beam is reflected by a splitter and focused by an objective with high numerical aperture. The Raman peaks variation across the edge is recorded by raster-scanning the sample with a piezo-stage. The acquisition time per pixel is of the order of few minutes. Gratings of 150 and 600 grooves/mm are used. The spatial resolution is $\sim$800nm. The power on the samples is well below 2mW, so that no shift, nor change in width of the Raman peaks is observed, thus ensuring no damage, nor heating.

All carbons show common features in their Raman spectra in the 800-2000 cm$^{-1}$ region, the so-called G and D peaks, which lie at around 1580 and 1350 cm$^{-1}$ respectively\cite{acfRS}. The G peak corresponds to the $E_{2g}$ phonon at the Brillouin zone center. The D peak is due to the breathing modes of sp$^2$ rings and requires a defect for its activation\cite{tuinstra,ramanprb}. It comes from TO phonons around the \textbf{K} point of the Brillouin zone\cite{tuinstra,ramanprb}, is active by double resonance (DR)\cite{bara,steffi2} and is strongly dispersive with excitation energy due to a Kohn Anomaly at \textbf{K}\cite{PiscanecPRL}. The activation process for the D peak is an inter-valley process as follows: i) a laser induced excitation of an electron/hole pair; ii) electron-phonon scattering with an
exchanged momentum $\textbf{q}\sim\textbf{K}$; iii) defect scattering; iv) electron/hole recombination. The D peak intensity is not related to the number of graphene layers, but only to the amount of disorder\cite{tuinstra,ramanprb}. Indeed, when moving from graphite to nanocrystalline graphite, the ratio between the intensity of D and G peak, I(D)/I(G), varies inversely with the size of the crystalline grain or inter-defect distance\cite{tuinstra,ramanprb}. DR can also happen as intra-valley process i.e. connecting two points belonging to the same cone around $\textbf{K}$ (or $\textbf{K}'$). This gives rise to the so-called D'peak, which can be seen around 1620 cm$^{-1}$ in defected graphite \cite{nemanich}. The 2D peak is the second order of the D peak. This is a single peak in monolayer graphene, whereas it splits in
four bands in bilayer graphene, reflecting the evolution of the band structure\cite{ACFRaman}. The 2D' peak is the second order of the D' peak. Since 2D and 2D' peaks originate from a Raman scattering process where momentum conservation is obtained by the participation of two phonons with opposite wavevectors ($\textbf{q}$ and $-\textbf{q}$), they do not require the presence of defects for their activation, and are thus always present.
\begin{figure}
\centerline{\includegraphics [width=70mm]{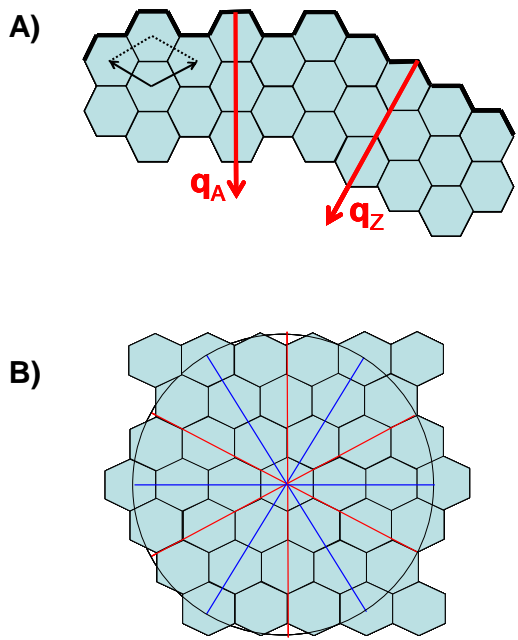}}
\caption{\label{Fig1} (color online)(a) The wavevector direction of electrons back-scattered by a zigzag, or armchair edge ($\textbf{d}_z$, $\textbf{d}_a$) is perpendicular to the edge; (b) Possible angles formed between two edges of graphene as a function of chirality. Red lines indicate armchair edges, while blue ones, the zigzag.}
\end{figure}
\begin{figure}
\centerline{\includegraphics[width=60mm] {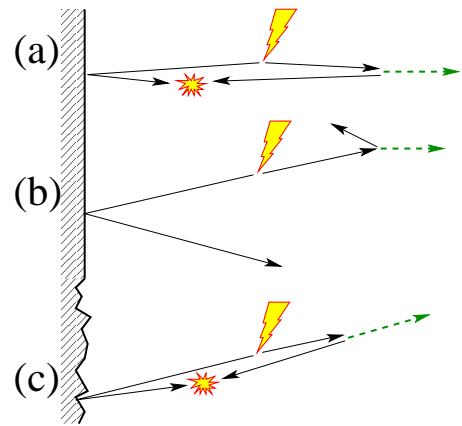}}
\caption{\label{Fig2New} (color online) Real space representation of the scattering process responsible for the D-peak at graphene edges. The lightning represents the incoming photon which generates the electron-hole pair; the solid black arrows, the quasi-classical trajectories of electron and hole; The dashed arrow, the emitted phonon; the flash, the radiative recombination of the electron-hole pair producing the scattered photon. (a)Backscattering off an ordered edge is possible only at normal incidence (up to a quantum diffraction correction $\sim\sqrt{\hbar\omega_{ph}/\epsilon}\ll{1}$). (b)For oblique incidence on an ordered edge the reflection is specular, so the electron and hole will not be able to meet at the same point. (c)For a disordered edge, backscattering is possible even at oblique incidence.
}\end{figure}

It is common for as prepared graphene not to have enough structural defects for the D peak to be seen\cite{ACFRaman}, indicative of the high crystallinity of graphene obtained by micro-mechanical cleavage. In this case, a D peak is only present at the edges\cite{ACFRaman}, since they act as defects, allowing elastic backscattering of electrons even in an otherwise defect-free sample. In their seminal work on graphite edges, Can\c{c}ado et al.\cite{cancado04, cancadorib} showed that the wavevector direction of the electrons back-scattered by an edge depends on edge chirality, as illustrated in Fig.\ref{Fig1}a. We now summarize and extend the results of Ref.\cite{cancado04, cancadorib}, by analyzing step by step the scattering process leading to the edge D-peak. This starts with the absorption of a photon of energy $\hbar\omega_L$ and the creation of an electron-hole pair ($e-h$). Assuming a perfect symmetry between valence and conduction bands, we can set the energies of the electron and hole, measured from the Dirac point, to be $\epsilon\approx\hbar\omega_L/2$. As discussed in Ref.\cite{BaskoBig}, energy conservation cannot be satisfied exactly in all elementary steps of the DR process responsible for the D peak, and the electron and hole energies have an uncertainty of $\sim \hbar\omega_{ph}$, where $\omega_{ph}$ is the phonon frequency. Still, as $\hbar\omega_{ph}\approx{0}.17\:\mbox{eV}$ is much smaller than the electron energy ($\epsilon\approx{1}.2\:\mbox{eV}$ for 514~nm excitation), we will speak about electron and hole energies. Since the photon momentum is negligible, the $e$ and $h$ wavevectors measured from $\textbf{K}$ have the modulus $k'=\epsilon/(\hbar v_F)$, where $v_F$ is the Fermi velocity. Following the approach of Ref.\cite{BaskoBig}, the photoexcited electron and hole can be viewed as wave packets of size $\sim{\hbar v_F}/{\epsilon}$, with real space velocity given by the slope of the reciprocal space band to which they belong. Therefore, assuming symmetric electronic bands, the photoexcited electron and hole move away with opposite velocities $\pm$\textbf{v} from the point where they have been created. An electron and hole traveling in opposite directions pose the problem of their radiative recombination, which is a key step of the total Raman scattering process. Indeed, as illustrated in Fig.~\ref{Fig2New}, the $e-h$ radiative recombination can occur only if the two particles, after undergoing a number of previous scattering events, find themselves at the same time in the same region of space (an area of size $\sim\hbar{v_F}/{\epsilon}$). In the case of the D-peak, the events that lead to the $e- h$ recombination are the inelastic scattering with a phonon and the elastic scattering with a defect. The radiative $e-h$ recombination requires momentum conservation. Thus, the negligible photons momentum imposes the recombining $e$, $h$ to have exactly opposite momentum. This, in turn, implies that in the DR process which makes the D peak Raman active, both phonon and defect scattering have to be back-scattering events\cite{BaskoBig}. Thus, the D peak activation by DR strongly depends on the nature of each scattering event. The overall size of the region where the process can take place is determined by the typical lifetime of the photoexcited electron-hole pair. For the D peak this size was recently measured by Can\c{c}ado et al.\cite{cancado08}.

\begin{figure}
\centerline{\includegraphics[width=90mm] {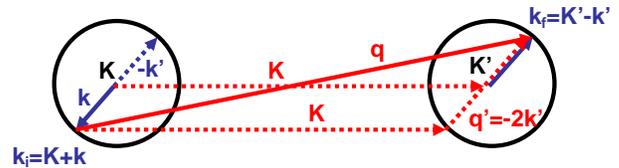}}
\caption{\label{Fig1New} (color online) Reciprocal space scheme of the $e-h$ backscattering condition. The circles represent isoenergy cuts on the Dirac cones. A phonon $\textbf{q}$ (solid red arrow) backscatters an electron from $\textbf{k}_i=\textbf{K}+\textbf{k}'$ to $\textbf{k}_f=\textbf{K}'-\textbf{k}'$. Since $\textbf{q}=\textbf{K}+\textbf{q}'$, back-scattering happens only if $\textbf{q}'=-2\textbf{k}'$.
}\end{figure}

We now analyze phonon and defect scattering. To activate the D-peak, phonon scattering has to satisfy two constraints: the first is that the phonon wavevector, \textbf{q}, must have one end on the Dirac cone around \textbf{K} and the
other on the cone around $\textbf{K}'$. The second is set by the back-scattering constraint. This implies that, if $\textbf{k}'$ and $\textbf{q}'$ are the electron and phonon wave-vectors measured from~\textbf{K}, we need $\textbf{q}'=-2\textbf{k}'$, as for Fig.~\ref{Fig1New}. This limits the subset of \textbf{q} satisfying the first condition. The defect scattering must also satisfy two constraints: momentum conservation with the phonon scattering, and back-scattering. Momentum conservation implies that the momentum exchanged by the defect elastic scattering has to be $\textbf{d}=-\textbf{q}$. This implies that the back-scattering condition $\textbf{d}=\textbf{K}'+2\textbf{k}'$ is automatically satisfied.

We now consider the specific case of D peak activation by edge scattering. It is crucial to distinguish between ordered (zigzag or amchair) and disordered edges. An important characteristic of a disordered edge is the length scale~$\xi$ at which it is disordered. The typical value of the momentum along the average edge direction, which can be transferred during a scattering event, is $d_\|\sim{1}/\xi$. For a perfect edge, the translational invariance along the edge imposes the momentum conservation along the edge, so $d_\|=0$, and the overall transferred momentum \textbf{d} is necessarily perpendicular to the edge.

In Fig.~\ref{Fig1}, electrons back-scattered by ordered armchair edges change their momentum by~\textbf{d}$_a$, while those by ordered zigzag edges, by~\textbf{d}$_z$. Fig. 1(a) shows that $\textbf{d}_a$ is directed along $\textbf{K}-\textbf{K}'$ (equivalent to $\textbf{K}'$), while \textbf{d}$_z$ along $\textbf{K}-\textbf{K}$ (here the equality is up to a reciprocal lattice vector). Since the D peak onset requires scattering between the two non-equivalent cones centered at $\textbf{K}$ and $\textbf{K}'$ (inter-valley scattering), the D peak cannot be produced by a perfect zigzag edge, while it should appear near a perfect armchair edge. This does not apply to the D' peak, where both \textbf{d}$_a$ and \textbf{d}$_z$ are compatible with the required intravalley scattering, Fig. 4b.

As discussed above, within the quasi-classical framework, electron and hole can be seen as wavepackets of size $\sim{\hbar v_F}/{\epsilon}$.
When such wavepacket reaches an edge, it is scattered. The edge region involved in this scattering has a linear extension ${\sim \hbar v_F}/{\epsilon}\sim 5-6\:\mbox{\AA}$ ($\sim$ 5 or 6 atoms). Thus, if $\xi\gg{\hbar v_F}/{\epsilon}$, the total exchanged momentum \textbf{d} will be perpendicular to the {\em local} edge direction, and the electron (hole) scattering against the edge resembles the elastic scattering of a ball against a wall, the corrections due to quantum diffraction being small ($\sim\hbar\omega_{ph}/\epsilon$ or $\sim\hbar{v}_F/\epsilon\xi$, whichever larger). In this case, the D peak is activated by the local armchair segments. Such edge may be called mixed or quasi-ordered.

On the contrary, if the edge is disordered on a shorter length scale, $\xi\sim{\hbar v_F}/{\epsilon}$, and both zigzag and armchair segments are present (or the edge is so disordered that no segments of any definite orientation can be identified) the total exchanged momentum \textbf{d} is randomly oriented. This means that (i) an electron (or hole) impinging perpendicularly to a disordered edge can be reflected in virtually any direction, and (ii)~an electron (or hole) moving at an oblique angle still has a chance to be reflected back and contribute to the D peak, as illustrated in Fig.~\ref{Fig2New}(c).
\begin{figure}
\centerline{\includegraphics [width=100mm]{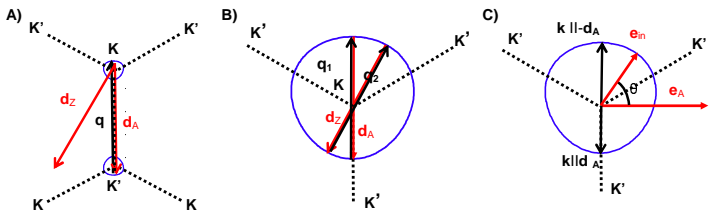}}
\caption{\label{Fig4}(a) (Color online) Schematic of inter-valley process: only the exchanged momentum from armchair edges $\textbf{d}_a$ can connect \textbf{K} and $\textbf{K}'$; b) Schematic of intra-valley process: both the exchanged momentum from armchair and zigzag edges $\textbf{d}_z$ and $\textbf{d}_a$ can connect points belonging to the same region around \textbf{K} ($\textbf{K}'$). The blue line shows the iso-energy contour, taking into account the trigonal warping,Ref.\cite{Kurti2002}; (c) Schematic of polarized Raman: when the incoming light is polarized $\textbf{e}_{in}$ the D intensity depends on $\theta$, the angle between the incident polarization and the armchair direction ($e_a$).}
\end{figure}

As we have seen, the D peak obtained from ordered and disordered edges originates from different directions of electron and hole motion. For an ordered armchair or mixed edge the electron and hole move perpendicularly to the edge (up to quantum diffraction corrections, allowing deviations by an angle $\sim\sqrt{\hbar\omega_{ph}/\epsilon}$). For a disordered edge the direction can be oblique (still, not completely arbitrary: if the momentum is directed along the edge, the electron/hole will never reach it). The directions are correlated with the polarization of the incident and scattered photons. Indeed, the probability of absorbing (emitting) a photon of polarization \textbf{e}$_{in}(\textbf{e}_{out})$ by creating (recombining) an electron-hole pair with velocities $\textbf{v},-\textbf{v}$ is proportional to $|\textbf{e} \times \textbf{v}|^2$. As a consequence, the maximum number of electron-hole pairs with wavevector parallel to the \textbf{d}$_a$ direction will be created when the incident radiation is polarized along the direction of an armchair edge \textbf{e}$_a$, and will decrease as $\cos^2\theta$, where $\theta$ is the angle formed by \textbf{e}$_a$ and \textbf{e}$_{in}$, Fig.4c. Thus, for ordered and mixed edges, the total intensity of the D peak is expected to be proportional to $\cos^4\theta$ or to $\sin^2\theta\cos^2\theta$ (up to quantum diffraction corrections) when the scattered photons are collected with a polarization parallel or orthogonal to $\textbf{e}_{in}$, respectively. For an unpolarized detection the intensity is given by their sum, $\cos^2\theta$. For an essentially disordered edge the process shown in Fig.~\ref{Fig2New}(c) is possible, and thus the intensity is finite even for light polarized perpendicularly to the edge. However, to compare this residual intensity to the quantum diffraction correction, a quantitative theory of electron scattering on a disordered edge is necessary, but this is not available at present.

To summarize, we distinguish four cases.
(i) If the edge is perfectly zigzag, the D peak is absent.
(ii) If the edge is perfectly armchair, the D peak has the strongest possible intensity. The armchair edge can be thus seen as a perfect defect. The excitation polarization dependence is $\cos^2\theta+O(\hbar\omega_{ph}/\epsilon)$, the correction being due to quantum diffraction.
(iii) If the edge is mixed, but still composed of locally perfect zigzag and armchair sections of typical size $\xi\gg\hbar{v}_F/\epsilon$, then only the armchair segments can contribute to DR. Thus, the D peak is expected to be smaller than that of a perfect armchair edge. Its polarization dependence is  $\cos^2\theta+O(\hbar{v}_F/\epsilon\xi)+O(\hbar\omega_{ph}/\epsilon)$ with respect to the direction of the armchair segments.
(iv) If the edge is essentially disordered (edges with different orientations, edges different from zigzag or armchair, pentagons, deformed rings and so on), then the intensity of the D peak should be even smaller than in case (iii). The minimum intensity measured as a function of polarization will have contributions due both to quantum diffraction and diffuse scattering at the edge.
\begin{figure}
\centerline{\includegraphics [width=90mm]{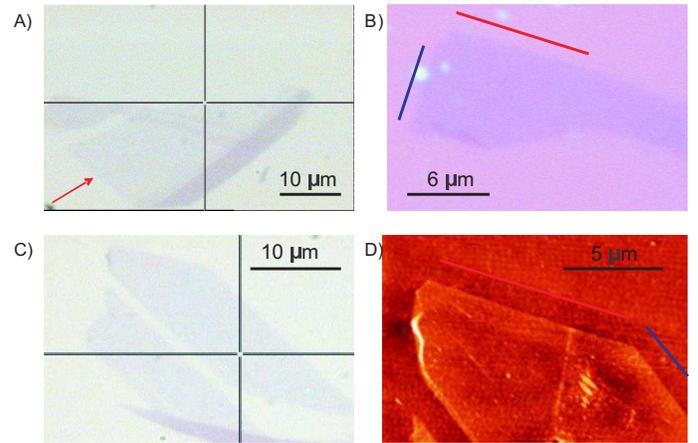}}
\caption{\label{Fig2} (a) (Color online) Optical micrograph of the flake used to study a single edge, indicated by the arrow; (b) flake containing two edges at an angle of 90$^\circ$; (c,d) Optical and AFM images of a flake with two edges at an angle of 150$^\circ$.
}\end{figure}

We now consider edge identification from optical micrographs. If we examine two edges in a graphene flake, and assume them to be ideal, their relative angle would depend only on their chirality, as shown in Fig. 2(b), where red lines identify armchair directions, and blue, zigzag directions. For example, if two edges form an angle of 120$^\circ$, both should have the same chirality. In contrast, an angle of 90$^\circ$ or 150$^\circ$ implies a change of chirality, Fig. 2(b). Note that,unless the orientation of the graphene layer is known \textit{a priori}, or one of the two edges assigned independently, absolute chirality assignment is not possible just based on their relative angle.
\begin{figure}
\centerline{\includegraphics [width=80mm]{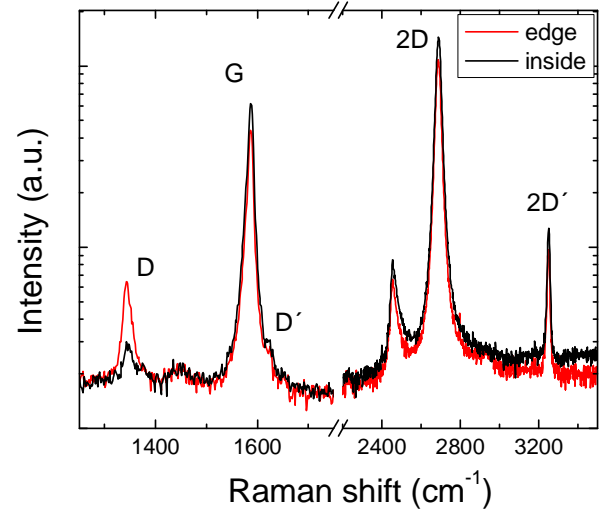}}
\caption{\label{Fig4} (a) (Color online) Raman spectra measured inside the sample (black) and at the edge (red) at 514 nm: a strong D peak is visible at the edge. Note that the peak at$\sim$1450cm$^{-1}$ is the third order of the silicon substrate\cite{ccdrm,silicon}.}\end{figure}

Fig.5(a,b,c) plot the optical micrographs of the three samples studied here. They are identified as single layer by Rayleigh\cite{CasiraghiNL} and Raman spectroscopy\cite{ACFRaman}. We select samples containing edges forming an angle of 90$^\circ$ and 150$^\circ$, as observed by optical and Atomic Force Microscopy (AFM), Fig.5(d). One can see that the edges appear uniform on a micron length scale.

We first consider the Raman spectra measured at 514nm for the edge indicated by the arrow in Fig. 5(a). The edge Raman spectrum, shown in red in Fig. 6, has a strong D peak (note the log scale of the y-axis used to enhance the smaller peaks). The peak at $\sim$1450 cm$^{-1}$ in Fig. 6 is not due to graphene, since it is visible on the substrate as well, but it is the third order Raman peak of the silicon substrate\cite{ccdrm,silicon,temple}. The D peak dispersion at the edge is $\sim$50cm$^{-1}$/eV, similar to the D peak inside graphite\cite{acfRS,pocsik, wang, matthews}. That of the 2D and 2D' peaks is$\sim$95 and$\sim$21cm$^{-1}$/eV, respectively, as Refs.\cite{Leandro,tan}.
\begin{figure}
\centerline{\includegraphics [width=85mm]{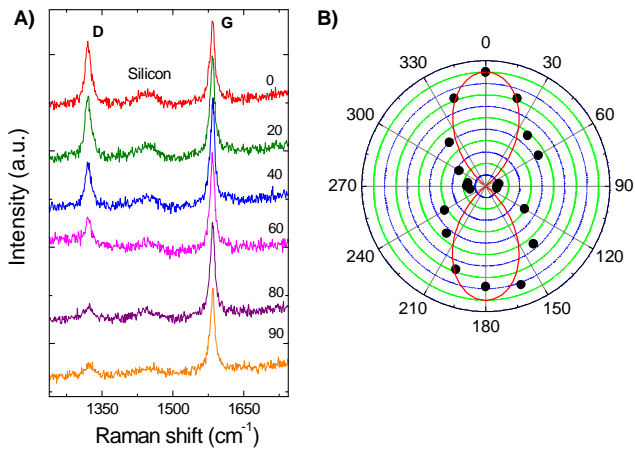}}
\caption{\label{Fig6} (Color online) (a) Raman spectra of one edge measured for different incident polarization at 633 nm. (b) I(D)/I(G) as a function of angle between light and edge. Note that I(D)/I(G) does not go to zero for perpendicular polarization. This indicates that the edges is not perfect.}\end{figure}

Fig. 7(a) plots the Raman spectra for different incident polarization. The D intensity strongly depends on the angle between incident polarization and edge: it is maximum when the polarization is parallel to the edge ($\theta=0^\circ,180^\circ$) and minimum when perpendicular ($\theta=90^\circ,270^\circ$). This agrees with what previously observed in polarized measurements at the edge of graphite and graphite ribbons\cite{cancado04, cancadorib}. Fig. 7(b) shows the dependence of I(D)/I(G) as a function of $\theta$ for the edge in Fig5(a). The D intensity is consistent with  a $\cos^2\theta$ dependence\cite{cancado04,cancadorib}, Fig. 7(b). However, it does not go to zero, as one would expect, for light polarized perpendicular to the edge, Fig. 7. The residual I(D)/I(G) area ratio at the edge is $\sim$0.3, much larger than in the bulk, where $\mbox{I(D)/I(G)}<0.1$. The ratio of the residual intensity I(D)$_{min}$ to the maximum intensity I(D)$_{max}$,
$\mbox{I(D)}_{min}/\mbox{I(D)}_{max}\sim0.1-0.2$, is compatible with the diffraction correction to the normal incidence condition. This would imply (iii), discussed above: mixed edge with alternating zigzag and armchair segments. Since the minimum is reached when the polarization is perpendicular to the {\em average} direction of the edge, this average direction should correspond to the armchair.
\begin{figure}
\centerline{\includegraphics [width=75mm]{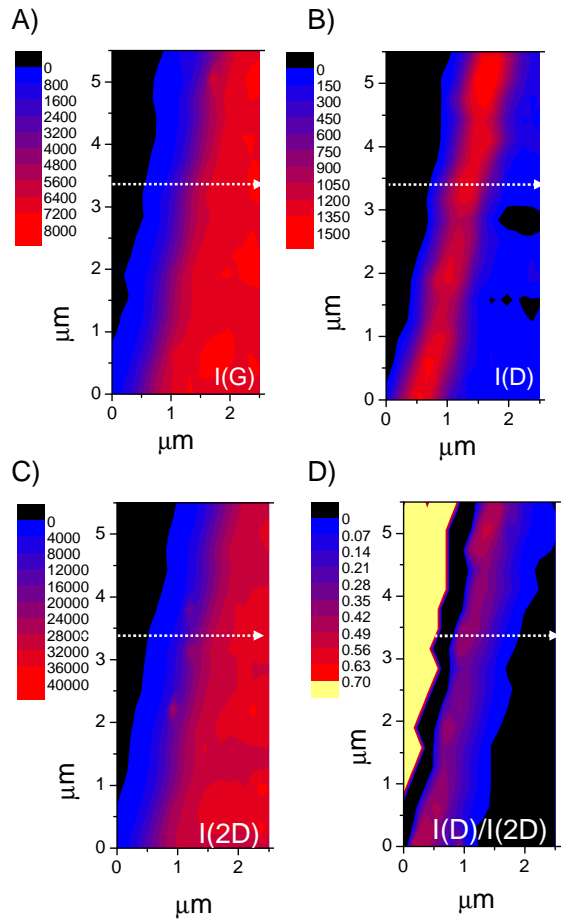}}
\caption{\label{Fig7} (a) Raman map of I(G) in proximity
of the graphene edge, shown in red in Figure 1; (b) Raman map of I(G); (c) Raman map of I(2D); (d) Raman map of I(D)/I(G).} \end{figure}

To understand how the intensity of the G, 2D and D peaks scales when crossing an edge, Raman mapping is performed with light polarized parallel to the edge direction. Figs. 8 show a map of I(G), I(D), I(2D), I(D)/I(G) across the edge. The D peak is localized at the edge, unlike the G peak, which increases, as expected, when moving from outside to inside the sample. Fig. 9 plots the profile of I(G), I(2D), I(D) and I(D)/I(G) across the edge. When going from outside to inside the flake, I(D) increases, reaches a maximum and then decreases. Fitting this variation with a gaussian, we get a width of $\sim$700 nm, comparable with our spatial resolution. On the other hand, I(G) decreases moving from inside to outside the flake. This is expected since the Raman intensity of the allowed peak is proportional to the volume of the sample. The same is observed for the 2D peak,Figs. 8,9.

The D peak behaves in a different way compared to the G and 2D, because its intensity is proportional to the amount of defects, which, neglecting structural disorder, can be assumed to be proportional to the edge length under the laser spot, as discussed above. Thus, the maximum I(D) should be measured when the diameter of the laser beam crosses the edge (dotted vertical line in Fig. 9). Here, I(G) is roughly half of I(G) inside the flake, Fig. 9. We define this position as the "edge". Thus, when we refer to any Raman parameter as measured at the edge, we mean measured at this position. This is needed to safely compare values of I(D)/I(G) measured on different edges or different points on the same edge.
\begin{figure}
\centerline{\includegraphics [width=60mm] {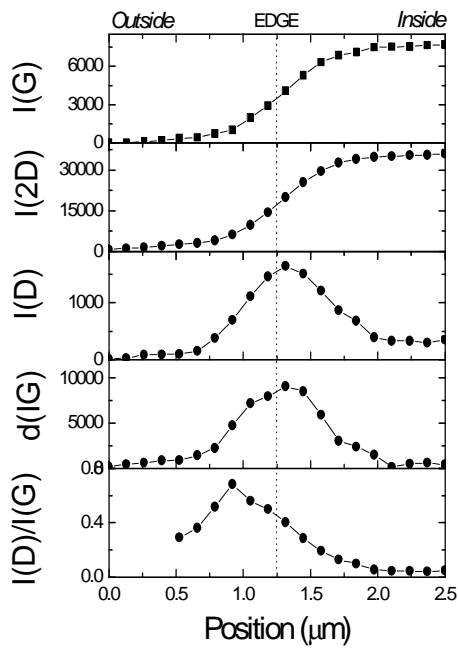}}
\caption{\label{Fig8} Profile variation of the intensity of G, 2D, D, derivative of the G peak and I(D)/I(G), measured along the white arrow in Figure 6.} \end{figure}

Fig. 10(a) shows the Raman spectra obtained when scanning across the edge. We detect a small red
shift of the G peak when the laser spot is almost outside the flake, accompanied by a FWHM decrease,
Fig. 10(b). This is not due to increasing disorder, since in this case we would expect FWHM(G) to
increase\cite{Casiraghi_condmat_2007}, but it could be related to a variation of
doping\cite{Casiraghi_condmat_2007,Pisana,DasCM}, i.e. this edge shows a slightly higher doping level
compared to the bulk. Fig. 10(b) also plots the position of the D peak. This decrease from 1348 to 1346 cm$^{-1}$ when moving from inside to the edge. This is very small, within the spectrometer resolution, and in opposite direction, compared to the G peak. Thus, stress is not the reason of this G
peak trend.
\begin{figure}
\centerline{\includegraphics [width=80mm] {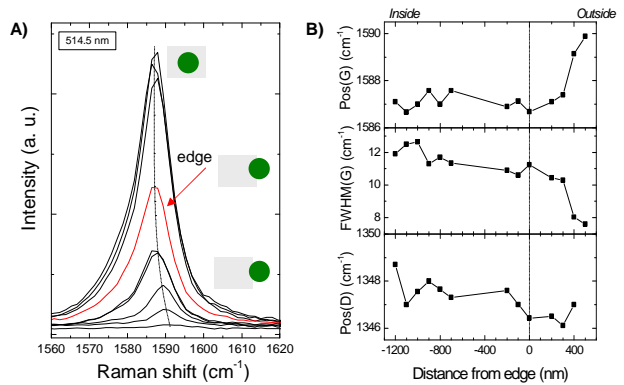}}
\caption{\label{Fig9} (a) Raman spectra obtained at 514 nm: the green circle represents the laser spot and the gray square is the graphene; (b) G, D peak positions and FWHM(G) as a function of the distance from the edge.}\end{figure}
\begin{figure}
\centerline{\includegraphics [width=80mm]{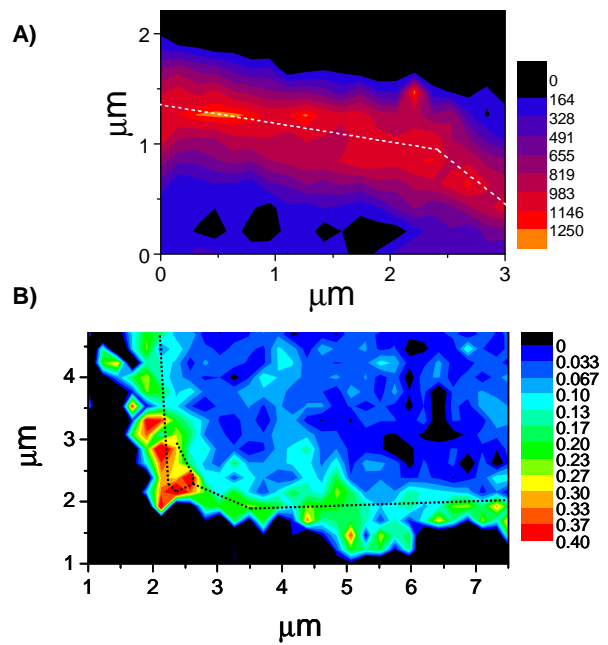}}
\caption{\label{Fig10} (a) Raman map of (ID) in proximity of two edges forming an angle of 150 degrees. I(D)
does not show any strong variation along the two edges. (b) Raman map of (ID)/I(G) in proximity of two edges forming an angle of 90 degrees. I(D)/I(G) is never null or at least comparable with what observed inside the flake on any edge.}
\end{figure}

Finally, we perform a Raman map on two edges forming angles of 90$^\circ$ and 150$^\circ$, Fig. 5(b,c). This angle implies that, if one edge is zigzag the other must be armchair or vice-versa, but the two edges cannot have the same chirality, as shown in Fig. 2(b). Raman mapping with circular polarization is used in order to avoid preferential enhancement of I(D). Figure 11(a,b) show the maps of I(D) and I(D)/I(G). No strong variation of I(D) is detected at the edges, Fig. 11(a); I(D)/I(G) is never null, Fig. 11(b), in contrast to what expected for ideal edges\cite{cancado04,cancadorib}. It is interesting to compare our data with those in Ref.\cite{cancado04} for two graphite edges with an angle of 150$^\circ$. Ref.\cite{cancado04} states that I(D) at the edge is never null due to some disorder at the edges. However, it is claimed that a small I(D)/I(G) can be taken as a signature of zigzag edges, while a large I(D)/I(G) indicates armchair edges. Here we find that none of the samples we analyzed have perfect edges on a microscopic scale, even though they look very smooth by optical microscopy. In particular, we never observe perfect zigzag edges since our measured I(D) is never null.

In conclusion, Raman spectroscopy is an ideal tool to probe graphene edges. The D to G peak ratio strongly depends on polarization, relative position of the laser spot with respect to the edge, and amount of edge disorder. In some samples, Raman mapping with circular polarization shows no significant dependence of the D peak intensity on the macroscopic edge orientation. This indicates that edges can be mixed and disordered, at least on the laser spot scale, even though they appear to follow well defined crystallographic directions at a larger scale.

We acknowledge L. C. Can\c{c}ado, P.H. Tan and S. Reich for useful discussions; A. Fasoli and V. Scardaci  for assistance with the samples. CC acknowledges the Oppenheimer and the Alexander von Humboldt Foundation; ACF the Royal Society and the European Research Council Grant NANOPOTS.


\begin{thebibliography}{99}

\bibitem{Nov306(2004)}K. S. Novoselov, A. K. Geim, S. V. Morozov, D. Jiang, Y. Zhang, S. V.Dubonos, I. V. Grigorieva, A. A. Firsov; Science, \textbf{306,} 666 (2004).

\bibitem{GeimRevNM}A. K. Geim, K. S. Novoselov; Nature Materials, \textbf{6,} 183 (2007).

\bibitem{CastroNetoRev}A. H. Castro Neto, F. Guinea, N. M. R. Peres, K. S. Novoselov, A. K. Geim; arXiv:0709.1163v1.

\bibitem{charlier} J. C. Charlier, P.C. Eklund, J. Zhu, A.C. Ferrari, Topics Appl. Phys.\textbf{111}, 673
(2008).

\bibitem{Nov438(2005)}K. S. Novoselov, A. K. Geim, S. V. Morozov, D. Jiang, M. I. Katsnelson, I. V.
    Grigorieva, S. V. Dubonos, and A. A. Firsov; Nature (London), \textbf{438,} 197 (2005).

\bibitem{Zhang438(2005)} Y. Zhang, Y.W. Tan, H. L. Stormer, and P. Kim; Nature (London), \textbf{438,} 201 (2005).

\bibitem{Nov315(2007)}K. S. Novoselov, Z. Jiang, Y. Zhang, S. V. Morozov, H. L. Stormer, U.
    Zeitler, J. C. Maan, G. S. Boebinger, P. Kim, and A. K. Geim; Science, \textbf{315,} 1379 (2007).

\bibitem{MorozovNov(2007)}S. V. Morozov, K. S. Novoselov, M. I. Katsnelson, F. Schedin, D. C. Elias, J. A. Jaszczak, A. K. Geim; Phys. Rev. Lett., \textbf{100,} 016602 (2008).

\bibitem{andrei} X. Du, I. Skachko, A. Barker, E. Y. Andrei, cond mat. arXiv:0802.2933

\bibitem{kimmob}  K. I. Bolotin, K. J. Sikes, J. Hone, H. L. Stormer, P. Kim  arXiv:0802.2389; K. I. Bolotin, K. J. Sikes, J. Hone, H. L. Stormer, P. Kim arXiv:0805.1830

\bibitem{Han}M. Y. Han, B. zylmaz, Y. Zhang, P. Kim; Phys. Rev. Lett., \textbf{98,} 206805 (2007).
\bibitem{Chen}Z. Chen, Y.M. Lin, M. Rooks, P. Avouris; Physica E, \textbf{40,} 228 (2007).

\bibitem{Zhang86}Y. Zhang, J. P. Small, W. V. Pontius, P. Kim; Appl. Phys. Lett., \textbf{86,} 073104 (2005).

\bibitem{Lemme}M. C. Lemme, T. J. Echtermeyer, M. Baus, and H. Kurz; IEEE El. Dev. Lett.,\textbf{28,} 4 (2007).

\bibitem{Casiraghi_condmat_2007}C. Casiraghi, S. Pisana, K. S. Novoselov, A. K. Geim, A. C.Ferrari; Appl. Phys. Lett., \textbf{91,} 233108 (2007).

\bibitem{ACFRaman}A. C. Ferrari, J. C. Meyer, V. Scardaci, C. Casiraghi, M. Lazzeri, F. Mauri, S. Piscanec, Da Jiang, K. S. Novoselov, S. Roth, A. K. Geim; Phys. Rev. Lett., \textbf{97,} 187401 (2006).

\bibitem {Leandro} L. M. Malard, J. Nilsson, D. C. Elias, J. C. Brant, F. Plentz, E. S. Alves, A. H. Castro Neto, M. A. Pimenta, Phys. Rev. B 76, 201401 (1999)

\bibitem{Pisana}S. Pisana, M. Lazzeri, C. Casiraghi, K. S. Novoselov, A. K. Geim, A. C. Ferrari, F. Mauri, Nat. Mat. \textbf{6,} 198 (2007).

\bibitem{PiscanecPRL}S. Piscanec, M. Lazzeri, F. Mauri, A. C. Ferrari, J. Robertson; Phys. Rev. Lett., \textbf{93,} 185503 (2004).

\bibitem{cancado08} L. G. Can\c{c}ado, R. Beams, L. Novotny Cond.Mat. 0802.2709

\bibitem{CasiraghiNL}C. Casiraghi, A. Hartschuh, E. Lidorikis, H. Qian, H. Harutyunyan, T. Gokus, K. S. Novoselov, A. C. Ferrari; Nano. Lett., \textbf{7,} 2711 (2007).

\bibitem{GeimAPL}P. Blake, E. W. Hill, A. H. Castro Neto, K. S. Novoselov, D. Jiang, R. Yang, T. J. Booth, A. K. Geim; Appl. Phys. Lett., \textbf{91,} 063124 (2007).
\bibitem{ACFRamanSSC} A. C. Ferrari; Solid State Comm., \textbf{143,} 47 (2007).

\bibitem{DasCM} A. Das, S. Pisana, S. Piscanec, B. Chakraborty, S. K. Saha, U. V. Waghmare, R. Yang, H. R. Krishnamurhthy, A. K. Geim, A. C. Ferrari, A. K. Sood Nature Nano. \textbf{3}, 210 (2008)

\bibitem{Lu75}X. Lu, H. Huang, N. Nemchuk, R. S. Ruoff; Appl. Phys. Lett., \textbf{75,} 193 (1999).

\bibitem{avourisPE} Z. H. Chen, Y-M. Lin, M. J. Rooks, P. Avouris, Physica E, \textbf{40}, 228 (2007).

\bibitem{Nakada1996}K. Nakada, M, Fujita, G. Dresselhaus, M. S. Dresselhaus; Phys. Rev. B, \textbf{54,} 17954 (1996).

\bibitem{Fujita1996}M. Fujita, K. Wakabayashi, K. Nakada, K. Kusakabe; J. Phys. Soc. Jap., \textbf{65,} 1920 (1996).

\bibitem{Miyamoto59}Y. Miyamoto, K. Nakada, M. Fujita; Phys. Rev. B, \textbf{59,} 9858 (1999).

\bibitem{Wakabayashi59}K. Wakabayashi, M. Fujita, H. Ajiki, M. Sigrist; Phys. Rev. B, \textbf{59,} 8271 (1999).

\bibitem{SonPRL}Y. W. Son, M. L. Cohen, S. G. Louie; Phys. Rev. Lett., \textbf{97,} 216803 (2006).

\bibitem{Pisani}L. Pisani, J. A. Chan, B. Montanari, N. M. Harrison; Phys. Rev. B, \textbf{75,} 064418 (2007).

\bibitem{Nakada1998}K. Nakada, M. Igami, M. Fujita; J.Phys. Soc.Jap., \textbf{67,} 2388 (1998).
\bibitem{Niimi}Y. Niimi, T. Matsui, H. Kambara, K. Tagami, M. Tsukada, H. Fukuyama; Phys. Rev. B, \textbf{73,} 085421 (2006).

\bibitem{Kobayashi2005}Y. Kobayashi, K. Fukui, T. Enoki, K. Kusakabe, Y. Kaburagi; Phys. Rev. B,\textbf{71,} 193406 (2005).

\bibitem{SolsGuineaCastroNeto}F. Sols, F. Guinea, A. H. Castro Neto; Phys. Rev. Lett., \textbf{99,}166803 (2007).

\bibitem{NiimiASS}Y. Niimi, T. Matsui, H. Kambara, K. Tagami, M. Tsukada, and H. Fukuyama; Appl. Surf. Sci., \textbf{241,} 43 (2005).

\bibitem{cervantes} F. Cervantes, S. Piscanec, G. Csanyi, A. C. Ferrari, Phys Rev. B, \textbf{77}, 165427 (2008)

\bibitem{acfRS} A. C. Ferrari, J. Robertson (eds), Raman spectroscopy in carbons: from nanotubes to diamond, Theme Issue, Phil. Trans. Roy. Soc. A \textbf{362}, 2267-2565 (2004).

\bibitem{tuinstra} F. Tuinstra, J. L. Koenig, J. Chem. Phys. 53, 1126 (1970).

\bibitem{ramanprb} A.C. Ferrari and J. Robertson Phys. Rev. B {\bf 61}, 14095 (2000); {\it ibid.} {\bf 64}, 075414 (2001).

\bibitem{bara} A. V. Baranov, A. N. Bekhterev, Y. S. Bobovich and V. I. Petrov, Opt. Spektrosk. \textbf{62}, 1036 (1987).

\bibitem{steffi2} C. Thomsen, S. Reich, Phys. Rev. Lett. 85, 5214 (2000).

\bibitem {nemanich} R. J. Nemanich, S. A. Solin, Phys. Rev. B 20, 392 (1979)

\bibitem{cancado04} L. G. Can\c{c}ado, M. A. Pimenta, B. R. A. Neves, M. S. S. Dantas, A. Jorio, Phys. Rev. Lett. 93, 247401 (2004).

\bibitem{cancadorib} L. G. Can\c{c}ado, M. A. Pimenta, B. R. A. Neves, G. Medeiros-Ribeiro, T. Enoki, Y. Kobayashi, K. Takai, K. Fukui, M. S. Dresselhaus, R. Saito, A. Jorio, Phys. Rev. Lett. 93, 047403 (2004).

\bibitem{BaskoBig} D. M. Basko, Phys. Rev. B \textbf{78}, 125418  (2008).

\bibitem{Kurti2002} J. Kurti et al., Phys. Rev. B \textbf{65}, 165433 (2002).

\bibitem{ccdrm} C. Casiraghi, A. C. Ferrari, J. Robertson, R. Ohr, M. Gradowski, D. Schneider, H. Hilgers, Diam. Rel. Mat.13, 1480 (2004).

\bibitem{silicon} W. P. Acker, B. Y., D. H. Leach, R. K. Chang, J. Appl. Phys. 64, 2263 (1988)

\bibitem{temple} P. A. Temple, C. E. Hathaway, Phys. Rev. B 7, 3685 (1973)

\bibitem {pocsik} I. P\'{o}ksik, M. Hundhausen, M. Ko\'{o}s, L. Ley, J. Non-cryt. Solids, 227, 53 (1998)

\bibitem {wang} Y. Wang, D. C. AlsmeR. L. McCreery, Chem. Mater. 2, 557 (1990)

\bibitem {matthews} M. J. Matthews, M. A. Pimenta, G. Dresselhaus, M. S. Dresselhaus, M. Endo, Phys. Rev. B 59, R6585 (1999)

\bibitem {tan} P. H. Tan, C. Y. Hu, J. Dong, W. C. Shen, B. F. Zhang, Phys. Rev. B 64, 214301 (2001); 64, 214301 (2001)

\end{thebibliography}
\end{document}